\documentclass[prl,twocolumn]{revtex4}
\usepackage{graphicx}
\usepackage{amsmath}

\begin{document}

\title{Optical supercavitation in soft-matter}

\author{C. Conti$^{1}$ and E. DelRe$^{2}$}
\affiliation{
$^1$CNR-ISC Institute for Complex Systems, Department of Physics, University Sapienza, Piazzale Aldo Moro 2, 00185, Rome (IT)\\
$^2$ Dep. of Electrical and Information Engineering, University of L'Aquila, 67100 L'Aquila, Italy}
\email{claudio.conti@roma1.infn.it}
\date{\today}

\begin{abstract}
We investigate theoretically, numerically and experimentally
nonlinear optical waves in an absorbing out-of-equilibrium 
colloidal material at the gelification transition.
At sufficiently high optical intensity, absorption is frustrated and light propagates into the medium.
The process is mediated by the formation of a matter-shock wave
due to optically induced thermodiffusion, and largely
resembles the mechanism of hydrodynamical supercavitation,
as it is accompanied by a {\it dynamic} phase-transition region between the beam and the absorbing material.
\end{abstract}


\maketitle
Nonlinear optical propagation in complex fluids is mediated by several and often competing effects, which are in many respect un-explored,
specifically in the presence of structural phase transitions.
Among the various involved phenomena, there are light induced 
re-orientational effects \cite{Tabiryan86,Khoo95,SimoniBook,Conti03,ContiPRL04}, 
electrostriction \cite{Palmer80,Ashkin82}, 
thermal \cite{ShenBook,Rotschild05,Dreischuh06,Ghofraniha07,Kartashov07}
and thermodiffusive phenomena \cite{ Ghofraniha07PRB, Ghofraniha09, Lamhot09}.

Theoretical and experimental works have recently put emphasis on electrostriction
and the corresponding solitary waves and beam self-trapping \cite{Conti05PRL,Conti06,Dholakia07,Anyfantakis08,Lee09,El-Ganainy09,Matuszewski09}. 
No evidence is reported on optical nonlinear waves in the presence
of typical complex processes, such as dynamic phase transitions and aging
(see, e.g, \cite{Dawson02, Trappe04, Cipelletti05, Sciortino05}).
Self-induced transparency with respect to scattering losses in a non absorbing medium,
in absence of structural transitions and thermal effects, was predicted in ref. \cite{Ganainy07}.

In this work we consider the nonlinear optical propagation of a micron-sized focused beam in a dye-doped, strongly absorbing, 
out-of-equilibrium nanoscale colloid, which
undergoes a dynamical arrest, i.e., a liquid-gel transition.
We show that, when this happens, the system behaviour changes from being 
strongly absorbitive, to a regime that allows beam propagation. This is accompained
by a shock-wave-front in the material that forms a depletion region of the absorbing molecules.
The process resembles cavitation and supercavitation hydrodynamic phenomena (see, e.g., \cite{BrennenBook}):
the shock-depleted region has vanishing concentration of colloidal beads and constitutes a non-arrested layer. 
This layer allows light propagation through the otherwise gelified and absorbing medium.
This happens in the very same way that a shock-induced gaseous region favors propulsion in  liquid supercavitation.
The effect is investigated theoretically, numerically and experimentally, and 
the findings underline the main role of time and space nonlocality \cite{DelRe03} in enhancing shock-phenomena \cite{Barsi07,Ghofraniha07}.

{\it Theory ---}
We consider a colloid able to absorb light and undergoing a dynamic transition, from a
liquid regime to a gelified state (high viscosity).
The material is formed by a host liquid (water) in which interacting nano-sized dye-doped (hence, light absorbing) colloidal beads are dispersed.
In the presence of strong light absorption, the leading phenomena are thermally induced, and re-orientational effects and electrostriction are negligible.
Specifically, we consider the variation of the refractive index due to the thermal density variation in the host liquid,
and the Soret effect, which is the diffusion of the colloidal beads due to a non-uniform temperature profile.
As the colloidal particles are driven by the light-induced thermal gradient, the local absorption and temperature change, so that
the real and the imaginary parts of the optical susceptibility are both affected.
Such a mechanism resembles the photorefractive beam nonlinearity \cite{Segev92}, where the optical beam generates a space-charge field that, through the
electro-optic effect, changes the refractive index. In our case, the role of the space-charge field is played by the temperature,
which drives the particle motion.

Non-equilibrium thermodynamics provide the general equations for modelling thermodiffusion.\cite{degrootbook}
Relevant variables are the temperature $T$ and the colloidal concentration $c$, governed by
the coupled fluxes ${\bf J}_T$ and ${\bf J}_c$
\begin{equation}
\partial_t c=\nabla\cdot {\bf J}_c\text{,}\;\partial_t T=\nabla\cdot {\bf J}_T+D_T \frac{\alpha_0(c)}{k_T}|A|^2\text{.}
\label{ec}
\end{equation}
The currents are given by
\begin{equation}
{\bf J}_T=D_T \nabla T+D_D \nabla c; {\bf J}_c=D_c \nabla c+S_T D_c c (1-c) \nabla T\text{,}
\label{currents}
\end{equation}
where $D_D$ is the Dafour coefficient ($D_D=0$ hereafter), $D_T$ is the thermal diffusion coefficient,
$D_c$ is the mass diffusion, $S_T$ is the Soret coefficient at the working temperature $T_0$,
$k_T$ is the thermal conductivity and $\alpha_0(c)$ the light absorption coefficient detailed below.
We neglect the temperature variation of the Soret coefficient.
The spatial distribution of the continuous wave optical field is given by the paraxial equation
\begin{equation}
i\frac{\partial A}{\partial z} +\frac{1}{2 k}\nabla_{xy}^2 A + k\frac{\Delta n(T)}{n} A=-i\alpha_0(c) A\text{,}
\label{paraxial}
\end{equation}
with $|A|^2$ the optical intensity, $k=2\pi n/\lambda$ the wavenumber, $n$ is the refractived index, $\lambda$ the wavelength.
$\Delta n$ is the optically induced refractive index perturbation, due to the temperature effect $\Delta n=\Delta T\partial n/\partial T$
with $\Delta T=T-T_0$  and $\alpha_0(c)=\alpha_B+c\partial\alpha/\partial c$ is the loss coeffient, with $\alpha_B$ the residual host liquid absorption in 
absence of the colloid $c=0$ ($\partial_T n<0$ and $\partial_c n>0$).

Introducing $\rho=c/c_0$ with $c_0$ the background concentration, $\theta=(T-T_0)/T_0$ the normalized temperature, and scaling spatial and temporal 
variables such that $\xi=x/w_0$, $\sigma=y/w_0$, $\zeta=z/z_0$, $\tau=t/t_0$, with 
$z_0=k w_0^2$ the diffraction length ($w_0$ is the beam waist) and $t_0=w_0^2/D_T$, Eqs.(\ref{ec}-\ref{paraxial}) are written as
\begin{eqnarray}
i\partial_\zeta \psi+\frac{1}{2}\nabla_{\perp}^2\psi-i \delta \theta \psi=-i\alpha (\rho+\rho_0)\psi \label{psieq}\\
\partial_\tau \theta = \nabla^2_{\epsilon} \theta + (\rho+\rho_0) |\psi|^2 \label{thetaeq};
\partial_\tau \rho= \eta \nabla^2_\epsilon \rho+ s \nabla_\epsilon \cdot(\rho \nabla_\epsilon \theta)\label{rhoeq}
\end{eqnarray}
with $\alpha=z_0 c_0 \partial_\alpha n$, $\delta=k |\partial_T n| T_0 z_0/n$,  $\nabla=\nabla_{\xi,\sigma}+{\hat{\boldsymbol\zeta}}\epsilon\partial_\zeta$, $\epsilon=1/k z_0\cong 10^{-2}$ ($\hat{\boldsymbol\zeta}$ is the unit vector of the $\zeta$ direction),
and $\psi=A/A_0$ with $A_0^2=k T_0/\alpha_0 w_0^2$.
In (\ref{thetaeq}), $\rho_0=t_0 \alpha_B A_0^2/k_T c_0 \partial_\alpha n$ accounts for the residual light absorption in the absence of the colloidal beads.
$\eta=D_c/D_T$ measures the strength of particles diffusion
in terms of heat diffusion ($D_T$), and decreases ($\eta<<1$) at the dynamics slowing down.
$s=S_T T_0 D_T/D_c$ is the normalized Soret coefficient.

As previously investigated \cite{Ghofraniha09}, the temporal scale for the onset of 
a stationary temperature profile is much faster than for thermal diffusion. 
This is a signature of soft-matter, where various independent temporal scales intervene.
In our case, $\rho$ is slowly varying with respect to $\theta$ and follows adiabatically; 
analogously $\psi$ follow both $\rho$ and $\theta$. 

$\theta$ is expressed in terms of $\rho$ and $|\psi|^2$,
by using Eq.(\ref{thetaeq}) in the stationary regime ($\partial_\tau\theta=0$),
and Eq.(\ref{rhoeq}) becomes ($\rho_0=0$ for simplicity)
\begin{equation}
\frac{\partial\rho}{\partial \tau}=\eta \nabla^2_{\epsilon}\rho+ s(\nabla_\epsilon \rho)\cdot(\nabla_\epsilon \theta)-s \rho^2 |\psi|^2
 \label{rhoeq1}\text{.}
\end{equation}
In the initial stage, $\rho\cong 1$, the temperature gradient term $\nabla_\epsilon \theta$ in
(\ref{rhoeq1}) is the heat flux generated by the absorbed
input beam (centered at $\zeta=0$ and $\xi=\sigma=0$), which is given by
$\theta= P/4\alpha \sqrt{(\epsilon r)^2+ \zeta^2}$
where $P=\int_0^\infty |\psi|^2 r dr$ is the normalized beam-power and $r=\sqrt{\xi^2+\sigma^2}$;
as obtained by the Green function for the heat equation. The beam decays much faster in space than $\theta$ and 
can be treated in (\ref{thetaeq}) as a Dirac delta  (highly nonlocal approximation for $\theta$). 
Correspondingly, in (\ref{rhoeq1}), 
the drift term is found to scale as $P/L^2$ where $L$ is the distance from the peak input optical intensity. 
Thus the temperature gradient induces a drift velocity of the density $\rho$, such that the density 
travels faster in proximity of the optical field, and slower in the surrounding region.
To address the resulting formation of a shock, we consider the one dimensional case ($\partial_\xi=\partial_\sigma=0$), such that $\nabla_\epsilon^2 \theta=\epsilon^2 \partial_\zeta^2\theta=-\rho |\psi|^2$ and $\epsilon^2 \partial_\zeta \theta=-\int_\zeta \rho I $. The heat source $\rho |\psi|^2$ after the depletion stage is 
a narrow localized regions (see Fig.\ref{figshock3} below), where the intensity $|\psi|^2$ decays as $\exp\left(-2 \rho \alpha \zeta\right)\cong \exp(-2 \alpha \zeta)$,
being $\rho\cong 1$. Correspondingly, $\epsilon^2 \partial_\zeta\theta\cong -\rho I_0/(2\alpha)$ and $I_0$ the input peak intensity.
The resulting equation for $\rho$ becomes 
\begin{equation}
\partial_{\tau}\rho+ \frac{s I_0}{2 \alpha}\rho \partial_\zeta \rho=\eta \epsilon^2 \partial_\zeta^2 \rho\text{,}
\label{Burgers}
\end{equation}
which is the Burgers equation predicting the formation of a shock front
in the $\zeta$ direction at the dynamic arrest ($\eta\rightarrow 0$) \cite{WhithamBook}. 
The term $\rho^2 |\psi|^2$ in (\ref{rhoeq1}) is exponentially small due to the light absorption and neglected.
Extending these arguments to the three-dimensional case leads to the prediction of a shock front with an semi-elliptical shape 
in the $(r,\zeta)$ plane, for a sufficiently high power 
and at the dynamic transition for the soft-colloidal matter.
We stress that for low enough power, and for a matter relaxation time comparable with the thermal one
(i.e., in the liquid phase, $\eta\cong1$), the shock is smoothed out by the ``dissipative'' term $\eta \epsilon^2 \partial_{\zeta}^2\rho$ in (\ref{Burgers}).
It is this light induced shock that produces a depletion of the absorbing material, which correspondingly becomes transparent
and allows light propagation. We stress that our model describes nonlinear optical propagation both in the liquid and in the arrested phase, which
are described by different values for $\eta$ (vanishing in the arrested phase and determined by the initial concentration $c_0$),
i.e., by different values of the matter diffusion coefficient $D_c$, which is the parameter mainly affected in the gelification process
(see also \cite{Ghofraniha07PRB, Ghofraniha09}).

{\it Spatio-temporal dynamics ---}
We numerically solved Eqs.(\ref{psieq}-\ref{rhoeq}) by using
a finite-difference time-domain predictor-corrector scheme for the temporal evolution 
of $\rho$ and $\theta$, and a pseudo-spectral beam-propagation scheme for $\psi$. 
At each instant the distribution of the field is calculated
from the distribution of $\rho$ and $\theta$. To reduce the computational effort,
we limited to a one-transverse dimension $\xi$ ($\partial_\sigma=0$). 
Parameters are chosen from realistic values.
For an input Gaussian beam (with amplitude $\hat A$ and unitary normalized waist),
we report representative simulations. For a large set of 
parameter values, the overal spatio-temporal dynamics are qualitatively the same, as described in what follows.

\begin{figure}
\includegraphics[width=0.48\textwidth]{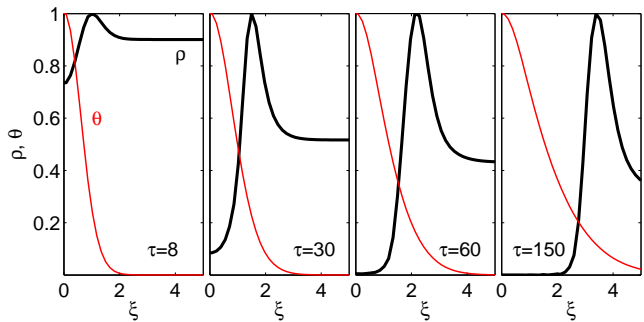}
\caption{(Color online) 
Initial stage of the formation of the cavitation. 
Thick line is material density ($\zeta=0$, $\xi>0$), thin line is temperature. Quantities are scaled to their
peak value ($\alpha=6$, $\delta=1$, $s=10$, $\eta=0.1$, $\rho_0=1$, $\hat A=0.1$). \label{figshock0}} \end{figure}

\begin{figure}
\includegraphics[width=0.48\textwidth]{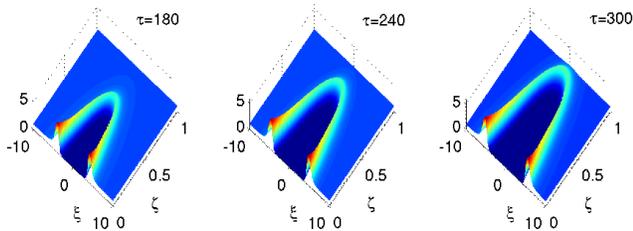}
\caption{(Color online) Density distribution $\rho$ (depletion layer) at different instants. Parameters as in Fig.\ref{figshock0}.
 \label{figshock1}} \end{figure}

\begin{figure}
\includegraphics[width=0.48\textwidth]{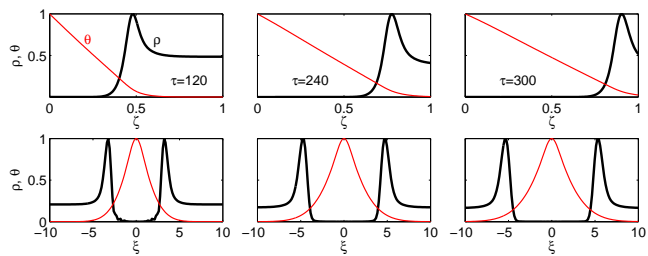}
\caption{(Color online)
Temperature (thin line) driven shock front in the longitudinal ($\xi=0$) and transverse ($\zeta=0$) sections 
(thick line is density). Parameters as in Fig.\ref{figshock0}.
\label{figshock2}} \end{figure}

After an initial stage for the formation of the temperature profile, 
the thermally driven diffusion starts to act on the $\rho$ field and induces a depletion region (where $\rho<<1$),
as in Fig.\ref{figshock0}.
As the density reaches the value $\rho=0$, the shock-front starts moving in the material. 
In Fig.\ref{figshock1} we show three snapshots of the density profile. 
Fig.\ref{figshock2} shows the longitudinal ($\xi=0$) and transverse ($\zeta=0$) shock profiles, superimposed
to the temperature, which unveil the traveling front driven by $\theta$.
Correspondingly, the material becomes transparent to the optical field,
whose propagation is forbidden in the linear regime by strong absorption.
These dynamics are obtained only at small values of $\eta$, while at higher values ($\eta\cong 1$) the diffusion of the
material smooths out and rapidly dissipates the shock front.
Fig.\ref{figshock3} shows the spatial beam intensity for three different instants. 
As the matter shock front advances, the beam enters the material that becomes progressively transparent.
We also report the fluorescence signal, given by $\rho |\psi|^2$, which appears as a propagating localized wave front and
is experimentally accessible. 
\begin{figure}
\includegraphics[width=0.50\textwidth]{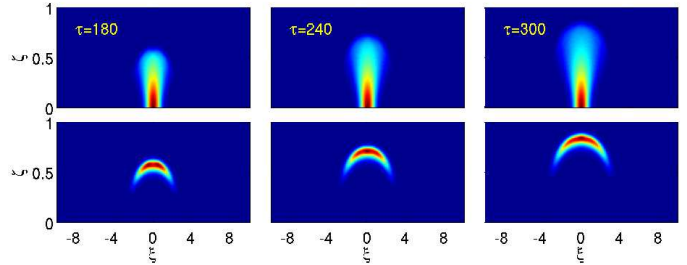}
\caption{(Color online)
Self-transparency mediated by the shock front. Top panels: beam intensity. 
Bottom: fluorescence signal $\rho |\psi|^2$. Parameters as in Fig.\ref{figshock0}.
\label{figshock3}} \end{figure}

{\it Experiments ---} In our experiments we use clay (Laponite) in water, doped by Rhodamine B (RhB),
a dye-doped colloidal solution displaying a gelification process \cite{Arbeloa98, Bonn99,Ruzicka04}.
The characterization of the nonlinear susceptibility was reported in \cite{Ghofraniha07PRB,Ghofraniha09}.
RhB molecules adsorb on the surface of the Laponite platelets (disks with diameter of the order of $25$nm), 
and the light absorption depends on $c$.
The sample (length $1$cm) is illuminated by a focused ($w_0=10\mu$m on the input facet) linearly polarized Gaussian beam
(frequency doubled continuos wave Nd:Yag $\lambda=$532nm, $n=1.3$),
obtained through a $40\times$ microscope objective.
Absorption losses are $\alpha_0\cong~5$mm$^{-1}$ (when $c=2$wt$\%$):
the sample appears red and is completely absorbing the $532$nm wavelength.
We studied two different conditions of concentration $c_0=1\%$ and $c_0=2\%$.
For $c_0=2$wt$\%$ the parameters previously defined are $z_0=1.5$mm, $\partial n/\partial T\cong 10^{-4}(^\circ$C$)^{-1}$, $T_0=300$K, 
$\epsilon\cong10^{-2}$,$\delta\cong 10$, $\alpha\cong 10$ ($D_c/D_T=\eta\cong0.1$).
The fluorescence signal is imaged by a microscope from the top of the sample.

\begin{figure}
\includegraphics[width=0.48\textwidth]{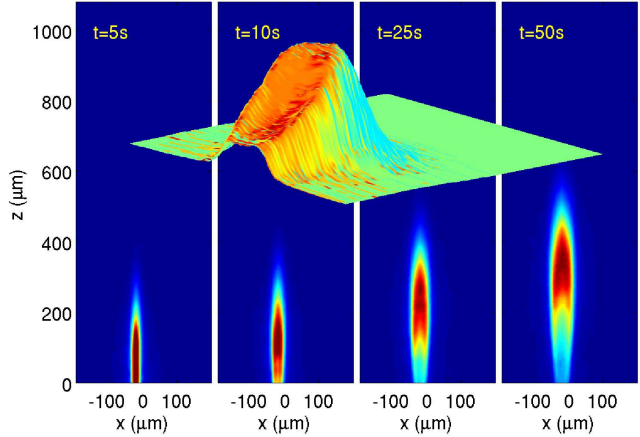}
\caption{(Color online)
Experimentally retrieved fluorescence signal due to the formation of a propagating beam in the material at three different instants.
The inset shows a three-dimensional representation at $t=100s$ (arbitrary scale).
\label{figexp1}} \end{figure}

At low concentration ($c_0=1$wt$\%$), the Laponite is far from the dynamic slowing down (liquid phase), 
and even at very high power levels $P_W\cong 1$W, no trasmission
is observed and the beam is rapidly absorbed along propagation. No significant time dynamics are 
observed. For $c_0=2\%$, at gelification, for power levels of $100$mW, a propagating light filament is observed,
tunnelling on a time scale of the order of tens of seconds (Fig. \ref{figexp1}).
The fluorescence signal is shown in fig. \ref{figexp1}.
The process is well rationalized by the theory and numerical experiments above.
The relaxation time after the shock phenomena is in 
general longer than that measured by dynamical light scattering or
z-scan \cite{Ghofraniha09}, which is of the
order of few seconds. This due to large structural modification (e.g.,enhanced packing) induced by the shock.

{\it Conclusions ---}
We reported on a novel regime of optical-propagation in soft-matter, 
mediated by many effects: diffraction,  absorption, temperature driven defocusing and thermophoresis.
The excitation of an hydrodynamic shock, resulting from a dynamically arrested phase, allows propagation inside an highly absorptive system.
This light-induced cavitation appears as a general process,
demonstrating how complex processes underlie nonlinear wave propagation in 
self-organized matter and lead to novel spatio-temporal effects.

We acknowledge support from CINECA.
The research leading to these results has received funding from the European
Research Council under the European Community's 
Seventh Framework Program (FP7/2007-2013)/ERC grant agreement n.201766.


\begin{thebibliography}{36}
\expandafter\ifx\csname natexlab\endcsname\relax\def\natexlab#1{#1}\fi
\expandafter\ifx\csname bibnamefont\endcsname\relax
  \def\bibnamefont#1{#1}\fi
\expandafter\ifx\csname bibfnamefont\endcsname\relax
  \def\bibfnamefont#1{#1}\fi
\expandafter\ifx\csname citenamefont\endcsname\relax
  \def\citenamefont#1{#1}\fi
\expandafter\ifx\csname url\endcsname\relax
  \def\url#1{\texttt{#1}}\fi
\expandafter\ifx\csname urlprefix\endcsname\relax\def\urlprefix{URL }\fi
\providecommand{\bibinfo}[2]{#2}
\providecommand{\eprint}[2][]{\url{#2}}

\bibitem[{\citenamefont{Tabiryan et~al.}(1986)\citenamefont{Tabiryan, Sukhov,
  and Zel'dovich}}]{Tabiryan86}
\bibinfo{author}{\bibfnamefont{N.~V.} \bibnamefont{Tabiryan}},
  \bibinfo{author}{\bibfnamefont{A.~V.} \bibnamefont{Sukhov}},
  \bibnamefont{and} \bibinfo{author}{\bibfnamefont{B.~Y.}
  \bibnamefont{Zel'dovich}}, \bibinfo{journal}{Mol. Cryst. Liq. Cryst.}
  \textbf{\bibinfo{volume}{136}}, \bibinfo{pages}{1} (\bibinfo{year}{1986}).

\bibitem[{\citenamefont{Khoo}(1995)}]{Khoo95}
\bibinfo{author}{\bibfnamefont{I.~C.} \bibnamefont{Khoo}},
  \emph{\bibinfo{title}{Liquid Crystals: Physical Properties and Nonlinear
  Optical Phenomena}} (\bibinfo{publisher}{Wiley}, \bibinfo{address}{New York},
  \bibinfo{year}{1995}).

\bibitem[{\citenamefont{Simoni}(1997)}]{SimoniBook}
\bibinfo{author}{\bibfnamefont{F.}~\bibnamefont{Simoni}},
  \emph{\bibinfo{title}{Nonlinear Optical Properties of Liquid Crystals}}
  (\bibinfo{publisher}{World Scientific}, \bibinfo{year}{1997}).

\bibitem[{\citenamefont{Conti et~al.}(2003)\citenamefont{Conti, Peccianti, and
  Assanto}}]{Conti03}
\bibinfo{author}{\bibfnamefont{C.}~\bibnamefont{Conti}},
  \bibinfo{author}{\bibfnamefont{M.}~\bibnamefont{Peccianti}},
  \bibnamefont{and} \bibinfo{author}{\bibfnamefont{G.}~\bibnamefont{Assanto}},
  \bibinfo{journal}{\prl} \textbf{\bibinfo{volume}{91}},
  \bibinfo{pages}{073901} (\bibinfo{year}{2003}).

\bibitem[{\citenamefont{Conti et~al.}(2004)\citenamefont{Conti, Peccianti, and
  Assanto}}]{ContiPRL04}
\bibinfo{author}{\bibfnamefont{C.}~\bibnamefont{Conti}},
  \bibinfo{author}{\bibfnamefont{M.}~\bibnamefont{Peccianti}},
  \bibnamefont{and} \bibinfo{author}{\bibfnamefont{G.}~\bibnamefont{Assanto}},
  \bibinfo{journal}{Phys. Rev. Lett.} \textbf{\bibinfo{volume}{92}},
  \bibinfo{pages}{113902} (\bibinfo{year}{2004}).

\bibitem[{\citenamefont{Palmer}(1980)}]{Palmer80}
\bibinfo{author}{\bibfnamefont{A.~J.} \bibnamefont{Palmer}},
  \bibinfo{journal}{Opt. Lett.} \textbf{\bibinfo{volume}{5}},
  \bibinfo{pages}{54} (\bibinfo{year}{1980}).

\bibitem[{\citenamefont{Ashkin et~al.}(1982)\citenamefont{Ashkin, Dziedzic, and
  Smith}}]{Ashkin82}
\bibinfo{author}{\bibfnamefont{A.}~\bibnamefont{Ashkin}},
  \bibinfo{author}{\bibfnamefont{J.~M.} \bibnamefont{Dziedzic}},
  \bibnamefont{and} \bibinfo{author}{\bibfnamefont{P.~W.} \bibnamefont{Smith}},
  \bibinfo{journal}{Opt. Lett.} \textbf{\bibinfo{volume}{7}},
  \bibinfo{pages}{276} (\bibinfo{year}{1982}).

\bibitem[{\citenamefont{Shen}(2002)}]{ShenBook}
\bibinfo{author}{\bibfnamefont{Y.~R.} \bibnamefont{Shen}},
  \emph{\bibinfo{title}{The Principles of Nonlinear Optics}}
  (\bibinfo{publisher}{Wiley-Interscience}, \bibinfo{year}{2002}).

\bibitem[{\citenamefont{Rotschild et~al.}(2005)\citenamefont{Rotschild, Cohen,
  Manela, Segev, and Carmon}}]{Rotschild05}
\bibinfo{author}{\bibfnamefont{C.}~\bibnamefont{Rotschild}},
  \bibinfo{author}{\bibfnamefont{O.}~\bibnamefont{Cohen}},
  \bibinfo{author}{\bibfnamefont{O.}~\bibnamefont{Manela}},
  \bibinfo{author}{\bibfnamefont{M.}~\bibnamefont{Segev}}, \bibnamefont{and}
  \bibinfo{author}{\bibfnamefont{T.}~\bibnamefont{Carmon}},
  \bibinfo{journal}{Phys. Rev. Lett.} \textbf{\bibinfo{volume}{95}},
  \bibinfo{pages}{213904} (\bibinfo{year}{2005}).

\bibitem[{\citenamefont{Dreischuh et~al.}(2006)\citenamefont{Dreischuh, Neshev,
  Petersen, Bang, and Krolikowski}}]{Dreischuh06}
\bibinfo{author}{\bibfnamefont{A.}~\bibnamefont{Dreischuh}},
  \bibinfo{author}{\bibfnamefont{D.~N.} \bibnamefont{Neshev}},
  \bibinfo{author}{\bibfnamefont{D.~E.} \bibnamefont{Petersen}},
  \bibinfo{author}{\bibfnamefont{O.}~\bibnamefont{Bang}}, \bibnamefont{and}
  \bibinfo{author}{\bibfnamefont{W.}~\bibnamefont{Krolikowski}},
  \bibinfo{journal}{Phys. Rev. Lett.} \textbf{\bibinfo{volume}{96}},
  \bibinfo{pages}{043901} (\bibinfo{year}{2006}).

\bibitem[{\citenamefont{Ghofraniha
  et~al.}(2007{\natexlab{a}})\citenamefont{Ghofraniha, Conti, Ruocco, and
  Trillo}}]{Ghofraniha07}
\bibinfo{author}{\bibfnamefont{N.}~\bibnamefont{Ghofraniha}},
  \bibinfo{author}{\bibfnamefont{C.}~\bibnamefont{Conti}},
  \bibinfo{author}{\bibfnamefont{G.}~\bibnamefont{Ruocco}}, \bibnamefont{and}
  \bibinfo{author}{\bibfnamefont{S.}~\bibnamefont{Trillo}},
  \bibinfo{journal}{Phys. Rev. Lett.} \textbf{\bibinfo{volume}{99}},
  \bibinfo{eid}{043903} (\bibinfo{year}{2007}{\natexlab{a}}).

\bibitem[{\citenamefont{Kartashov and Torner}(2007)}]{Kartashov07}
\bibinfo{author}{\bibfnamefont{Y.~V.} \bibnamefont{Kartashov}}
  \bibnamefont{and} \bibinfo{author}{\bibfnamefont{L.}~\bibnamefont{Torner}},
  \bibinfo{journal}{Opt. Lett.} \textbf{\bibinfo{volume}{32}},
  \bibinfo{pages}{946} (\bibinfo{year}{2007}).

\bibitem[{\citenamefont{Ghofraniha
  et~al.}(2007{\natexlab{b}})\citenamefont{Ghofraniha, Conti, and
  Ruocco}}]{Ghofraniha07PRB}
\bibinfo{author}{\bibfnamefont{N.}~\bibnamefont{Ghofraniha}},
  \bibinfo{author}{\bibfnamefont{C.}~\bibnamefont{Conti}}, \bibnamefont{and}
  \bibinfo{author}{\bibfnamefont{G.}~\bibnamefont{Ruocco}},
  \bibinfo{journal}{Phys. Rev. B} \textbf{\bibinfo{volume}{75}},
  \bibinfo{pages}{224203} (\bibinfo{year}{2007}{\natexlab{b}}).

\bibitem[{\citenamefont{Ghofraniha et~al.}(2009)\citenamefont{Ghofraniha,
  Conti, Ruocco, and Zamponi}}]{Ghofraniha09}
\bibinfo{author}{\bibfnamefont{N.}~\bibnamefont{Ghofraniha}},
  \bibinfo{author}{\bibfnamefont{C.}~\bibnamefont{Conti}},
  \bibinfo{author}{\bibfnamefont{G.}~\bibnamefont{Ruocco}}, \bibnamefont{and}
  \bibinfo{author}{\bibfnamefont{F.}~\bibnamefont{Zamponi}},
  \bibinfo{journal}{Phys. Rev. Lett.} \textbf{\bibinfo{volume}{102}},
  \bibinfo{eid}{038303} (\bibinfo{year}{2009}).

\bibitem[{\citenamefont{Lamhot et~al.}(2009)\citenamefont{Lamhot, Barak,
  Rotschild, Segev, Saraf, Lifshitz, Marmur, El-Ganainy, and
  Christodoulides}}]{Lamhot09}
\bibinfo{author}{\bibfnamefont{Y.}~\bibnamefont{Lamhot}},
  \bibinfo{author}{\bibfnamefont{A.}~\bibnamefont{Barak}},
  \bibinfo{author}{\bibfnamefont{C.}~\bibnamefont{Rotschild}},
  \bibinfo{author}{\bibfnamefont{M.}~\bibnamefont{Segev}},
  \bibinfo{author}{\bibfnamefont{M.}~\bibnamefont{Saraf}},
  \bibinfo{author}{\bibfnamefont{E.}~\bibnamefont{Lifshitz}},
  \bibinfo{author}{\bibfnamefont{A.}~\bibnamefont{Marmur}},
  \bibinfo{author}{\bibfnamefont{R.}~\bibnamefont{El-Ganainy}},
  \bibnamefont{and} \bibinfo{author}{\bibfnamefont{D.~N.}
  \bibnamefont{Christodoulides}}, \bibinfo{journal}{Phys. Rev. Lett.}
  \textbf{\bibinfo{volume}{103}}, \bibinfo{pages}{264503}
  (\bibinfo{year}{2009}).

\bibitem[{\citenamefont{Conti et~al.}(2005)\citenamefont{Conti, Ruocco, and
  Trillo}}]{Conti05PRL}
\bibinfo{author}{\bibfnamefont{C.}~\bibnamefont{Conti}},
  \bibinfo{author}{\bibfnamefont{G.}~\bibnamefont{Ruocco}}, \bibnamefont{and}
  \bibinfo{author}{\bibfnamefont{S.}~\bibnamefont{Trillo}},
  \bibinfo{journal}{\prl} \textbf{\bibinfo{volume}{95}},
  \bibinfo{pages}{183902} (\bibinfo{year}{2005}).

\bibitem[{\citenamefont{Conti et~al.}(2006)\citenamefont{Conti, Ghofraniha,
  Ruocco, and Trillo}}]{Conti06}
\bibinfo{author}{\bibfnamefont{C.}~\bibnamefont{Conti}},
  \bibinfo{author}{\bibfnamefont{N.}~\bibnamefont{Ghofraniha}},
  \bibinfo{author}{\bibfnamefont{G.}~\bibnamefont{Ruocco}}, \bibnamefont{and}
  \bibinfo{author}{\bibfnamefont{S.}~\bibnamefont{Trillo}},
  \bibinfo{journal}{Phys. Rev. Lett.} \textbf{\bibinfo{volume}{97}},
  \bibinfo{eid}{123903} (\bibinfo{year}{2006}).

\bibitem[{\citenamefont{Reece et~al.}(2007)\citenamefont{Reece, Wright, and
  Dholakia}}]{Dholakia07}
\bibinfo{author}{\bibfnamefont{P.~J.} \bibnamefont{Reece}},
  \bibinfo{author}{\bibfnamefont{E.~M.} \bibnamefont{Wright}},
  \bibnamefont{and} \bibinfo{author}{\bibfnamefont{K.}~\bibnamefont{Dholakia}},
  \bibinfo{journal}{Phys. Rev. Lett.} \textbf{\bibinfo{volume}{98}},
  \bibinfo{pages}{203902} (\bibinfo{year}{2007}).

\bibitem[{\citenamefont{Anyfantakis et~al.}(2008)\citenamefont{Anyfantakis,
  Loppinet, Fytas, and Pispas}}]{Anyfantakis08}
\bibinfo{author}{\bibfnamefont{M.}~\bibnamefont{Anyfantakis}},
  \bibinfo{author}{\bibfnamefont{B.}~\bibnamefont{Loppinet}},
  \bibinfo{author}{\bibfnamefont{G.}~\bibnamefont{Fytas}}, \bibnamefont{and}
  \bibinfo{author}{\bibfnamefont{S.}~\bibnamefont{Pispas}},
  \bibinfo{journal}{Opt. Lett.} \textbf{\bibinfo{volume}{33}},
  \bibinfo{pages}{2839} (\bibinfo{year}{2008}).

\bibitem[{\citenamefont{Lee et~al.}(2009)\citenamefont{Lee, El-Ganainy,
  Christodoulides, Dholakia, and Wright}}]{Lee09}
\bibinfo{author}{\bibfnamefont{W.~M.} \bibnamefont{Lee}},
  \bibinfo{author}{\bibfnamefont{R.}~\bibnamefont{El-Ganainy}},
  \bibinfo{author}{\bibfnamefont{D.~N.} \bibnamefont{Christodoulides}},
  \bibinfo{author}{\bibfnamefont{K.}~\bibnamefont{Dholakia}}, \bibnamefont{and}
  \bibinfo{author}{\bibfnamefont{E.~M.} \bibnamefont{Wright}},
  \bibinfo{journal}{Opt. Express} \textbf{\bibinfo{volume}{17}},
  \bibinfo{pages}{10277} (\bibinfo{year}{2009}).

\bibitem[{\citenamefont{El-Ganainy et~al.}(2009)\citenamefont{El-Ganainy,
  Christodoulides, Wright, Lee, and Dholakia}}]{El-Ganainy09}
\bibinfo{author}{\bibfnamefont{R.}~\bibnamefont{El-Ganainy}},
  \bibinfo{author}{\bibfnamefont{D.~N.} \bibnamefont{Christodoulides}},
  \bibinfo{author}{\bibfnamefont{E.~M.} \bibnamefont{Wright}},
  \bibinfo{author}{\bibfnamefont{W.~M.} \bibnamefont{Lee}}, \bibnamefont{and}
  \bibinfo{author}{\bibfnamefont{K.}~\bibnamefont{Dholakia}},
  \bibinfo{journal}{Phys. Rev. A} \textbf{\bibinfo{volume}{80}},
  \bibinfo{pages}{053805} (\bibinfo{year}{2009}).

\bibitem[{\citenamefont{Matuszewski et~al.}(2009)\citenamefont{Matuszewski,
  Krolikowski, and Kivshar}}]{Matuszewski09}
\bibinfo{author}{\bibfnamefont{M.}~\bibnamefont{Matuszewski}},
  \bibinfo{author}{\bibfnamefont{W.}~\bibnamefont{Krolikowski}},
  \bibnamefont{and} \bibinfo{author}{\bibfnamefont{Y.~S.}
  \bibnamefont{Kivshar}}, \bibinfo{journal}{Phys. Rev. A}
  \textbf{\bibinfo{volume}{79}}, \bibinfo{pages}{023814}
  (\bibinfo{year}{2009}).

\bibitem[{\citenamefont{Dawson}(2002)}]{Dawson02}
\bibinfo{author}{\bibfnamefont{K.~A.} \bibnamefont{Dawson}},
  \bibinfo{journal}{Curr. Opin. Colloid Interface Sci.}
  \textbf{\bibinfo{volume}{7}}, \bibinfo{pages}{218} (\bibinfo{year}{2002}).

\bibitem[{\citenamefont{Trappe and Sandkuhler}(2004)}]{Trappe04}
\bibinfo{author}{\bibfnamefont{V.}~\bibnamefont{Trappe}} \bibnamefont{and}
  \bibinfo{author}{\bibfnamefont{P.}~\bibnamefont{Sandkuhler}},
  \bibinfo{journal}{Curr. Opin. Colloid Interface Sci.}
  \textbf{\bibinfo{volume}{8}}, \bibinfo{pages}{494} (\bibinfo{year}{2004}).

\bibitem[{\citenamefont{Cipelletti and Ramos}(2005)}]{Cipelletti05}
\bibinfo{author}{\bibfnamefont{L.}~\bibnamefont{Cipelletti}} \bibnamefont{and}
  \bibinfo{author}{\bibfnamefont{L.}~\bibnamefont{Ramos}}, \bibinfo{journal}{J.
  Phys. : Condens. Matter} \textbf{\bibinfo{volume}{17}}, \bibinfo{pages}{R253}
  (\bibinfo{year}{2005}).

\bibitem[{\citenamefont{Sciortino and Tartaglia}(2005)}]{Sciortino05}
\bibinfo{author}{\bibfnamefont{F.}~\bibnamefont{Sciortino}} \bibnamefont{and}
  \bibinfo{author}{\bibfnamefont{P.}~\bibnamefont{Tartaglia}},
  \bibinfo{journal}{Advances in Physics} \textbf{\bibinfo{volume}{54}},
  \bibinfo{pages}{471} (\bibinfo{year}{2005}).

\bibitem[{\citenamefont{El-Ganainy et~al.}(2007)\citenamefont{El-Ganainy,
  Christodoulides, Rotschild, and Segev}}]{Ganainy07}
\bibinfo{author}{\bibfnamefont{R.}~\bibnamefont{El-Ganainy}},
  \bibinfo{author}{\bibfnamefont{D.~N.} \bibnamefont{Christodoulides}},
  \bibinfo{author}{\bibfnamefont{C.}~\bibnamefont{Rotschild}},
  \bibnamefont{and} \bibinfo{author}{\bibfnamefont{M.}~\bibnamefont{Segev}},
  \bibinfo{journal}{Opt. Express} \textbf{\bibinfo{volume}{15}},
  \bibinfo{pages}{10207} (\bibinfo{year}{2007}).

\bibitem[{\citenamefont{Brennen}(1995)}]{BrennenBook}
\bibinfo{author}{\bibfnamefont{C.}~\bibnamefont{Brennen}},
  \emph{\bibinfo{title}{Cavitation and Bubble Dynamics}}
  (\bibinfo{publisher}{Oxford University Press}, \bibinfo{address}{Oxford,
  England}, \bibinfo{year}{1995}).

\bibitem[{\citenamefont{Dari-Salisburgo
  et~al.}(2003)\citenamefont{Dari-Salisburgo, DelRe, and Palange}}]{DelRe03}
\bibinfo{author}{\bibfnamefont{C.}~\bibnamefont{Dari-Salisburgo}},
  \bibinfo{author}{\bibfnamefont{E.}~\bibnamefont{DelRe}}, \bibnamefont{and}
  \bibinfo{author}{\bibfnamefont{E.}~\bibnamefont{Palange}},
  \bibinfo{journal}{Phys. Rev. Lett.} \textbf{\bibinfo{volume}{91}},
  \bibinfo{pages}{263903} (\bibinfo{year}{2003}).

\bibitem[{\citenamefont{Barsi et~al.}(2007)\citenamefont{Barsi, Wan, Sun, and
  Fleischer}}]{Barsi07}
\bibinfo{author}{\bibfnamefont{C.}~\bibnamefont{Barsi}},
  \bibinfo{author}{\bibfnamefont{W.}~\bibnamefont{Wan}},
  \bibinfo{author}{\bibfnamefont{C.}~\bibnamefont{Sun}}, \bibnamefont{and}
  \bibinfo{author}{\bibfnamefont{J.~W.} \bibnamefont{Fleischer}},
  \bibinfo{journal}{Opt. Lett.} \textbf{\bibinfo{volume}{32}},
  \bibinfo{pages}{2930} (\bibinfo{year}{2007}).

\bibitem[{\citenamefont{Segev et~al.}(1992)\citenamefont{Segev, Crosignani,
  Yariv, and Fischer}}]{Segev92}
\bibinfo{author}{\bibfnamefont{M.}~\bibnamefont{Segev}},
  \bibinfo{author}{\bibfnamefont{B.}~\bibnamefont{Crosignani}},
  \bibinfo{author}{\bibfnamefont{A.}~\bibnamefont{Yariv}}, \bibnamefont{and}
  \bibinfo{author}{\bibfnamefont{B.}~\bibnamefont{Fischer}},
  \bibinfo{journal}{Phys. Rev. Lett.} \textbf{\bibinfo{volume}{68}},
  \bibinfo{pages}{923} (\bibinfo{year}{1992}).

\bibitem[{\citenamefont{{de Groot} and Mazur}(1984)}]{degrootbook}
\bibinfo{author}{\bibfnamefont{S.~R.} \bibnamefont{{de Groot}}}
  \bibnamefont{and} \bibinfo{author}{\bibfnamefont{P.}~\bibnamefont{Mazur}},
  \emph{\bibinfo{title}{Non-equilibrium thermodynamics}}
  (\bibinfo{publisher}{Dover, New York}, \bibinfo{year}{1984}).

\bibitem[{\citenamefont{Whitham}(1999)}]{WhithamBook}
\bibinfo{author}{\bibfnamefont{G.~B.} \bibnamefont{Whitham}},
  \emph{\bibinfo{title}{Linear and Nonlinear Waves}}
  (\bibinfo{publisher}{Wiley}, \bibinfo{address}{New York},
  \bibinfo{year}{1999}).

\bibitem[{\citenamefont{{L{\'o}pez Arbeloa}
  et~al.}(1998)\citenamefont{{L{\'o}pez Arbeloa}, {Herr{\'a}n Mart{\'i}nez},
  {L{\'o}pez Arbeloa}, and {L{\'o}pez Arbeloa}}}]{Arbeloa98}
\bibinfo{author}{\bibfnamefont{F.}~\bibnamefont{{L{\'o}pez Arbeloa}}},
  \bibinfo{author}{\bibfnamefont{J.~M.} \bibnamefont{{Herr{\'a}n
  Mart{\'i}nez}}}, \bibinfo{author}{\bibfnamefont{T.}~\bibnamefont{{L{\'o}pez
  Arbeloa}}}, \bibnamefont{and}
  \bibinfo{author}{\bibfnamefont{I.}~\bibnamefont{{L{\'o}pez Arbeloa}}},
  \bibinfo{journal}{Langmuir} \textbf{\bibinfo{volume}{14}},
  \bibinfo{pages}{4566} (\bibinfo{year}{1998}).

\bibitem[{\citenamefont{Bonn et~al.}(1999)\citenamefont{Bonn, Tanaka, Wegdam,
  Kellay, and Meunier}}]{Bonn99}
\bibinfo{author}{\bibfnamefont{D.}~\bibnamefont{Bonn}},
  \bibinfo{author}{\bibfnamefont{H.}~\bibnamefont{Tanaka}},
  \bibinfo{author}{\bibfnamefont{G.}~\bibnamefont{Wegdam}},
  \bibinfo{author}{\bibfnamefont{H.}~\bibnamefont{Kellay}}, \bibnamefont{and}
  \bibinfo{author}{\bibfnamefont{J.}~\bibnamefont{Meunier}},
  \bibinfo{journal}{Europhys. Lett.} \textbf{\bibinfo{volume}{45}},
  \bibinfo{pages}{52} (\bibinfo{year}{1999}).

\bibitem[{\citenamefont{Ruzicka et~al.}(2004)\citenamefont{Ruzicka, Zulian, and
  Ruocco}}]{Ruzicka04}
\bibinfo{author}{\bibfnamefont{B.}~\bibnamefont{Ruzicka}},
  \bibinfo{author}{\bibfnamefont{L.}~\bibnamefont{Zulian}}, \bibnamefont{and}
  \bibinfo{author}{\bibfnamefont{G.}~\bibnamefont{Ruocco}},
  \bibinfo{journal}{Phys. Rev. Lett.} \textbf{\bibinfo{volume}{93}},
  \bibinfo{pages}{258301} (\bibinfo{year}{2004}).

\end{thebibliography}

\end{document}